\documentclass{article}

\usepackage[utf8]{inputenc}
\usepackage[T1]{fontenc}
\usepackage{hyperref}
\usepackage{url}
\usepackage{booktabs}
\usepackage{amsfonts}
\usepackage{nicefrac}
\usepackage{microtype}
\usepackage{graphicx}
\usepackage{tabularx}
\usepackage{array}
\usepackage{enumitem}
\usepackage{orcidlink}

\title{From Awareness to Action: How Developers Engage with Accessibility Innovation in LLM-Assisted Development}

\author{
  Thayssa Águila da Rocha\orcidlink{0000-0001-8026-4003} \\
  Universidade Federal do Pará\\
  Zup Innovation\\
  \texttt{thayssa.rocha@icen.ufpa.br} \\
  \and
  Luciane Silva\orcidlink{0009-0000-2827-7886} \\
  Universidade Federal de Uberlândia\\
  \texttt{lucianefs@ufu.br} \\
  \and
  Ana Duarte \\
  Zup Innovation\\
  \texttt{ana.duarte@zup.com.br} \\
  \and
  Marcelle Pereira Mota\orcidlink{0000-0001-9226-9020} \\
  Universidade Federal do Pará\\
  \texttt{mpmota@ufpa.br} \\
  \and
  Gustavo Pinto\orcidlink{0000-0001-7598-2799} \\
  Universidade Federal do Pará\\
  \texttt{gpinto@ufpa.br} \\
}

\begin{document}

\maketitle

\begin{abstract}
Developers often struggle to design truly accessible digital solutions in corporate environments. In these environments, accessibility is usually treated as a compliance requirement rather than an innovation opportunity. By analyzing 14 LLM-based accessibility project proposals and focus group discussions with 9 participants at a Brazilian tech company, we found that inclusive innovation can emerge particularly when initiatives are led by People with Disabilities (PWD) themselves. If organizations adopt similar participatory approaches, accessibility would evolve from an afterthought into a driving force for technological excellence and cultural transformation.
\end{abstract}

\section{Introduction}

Accessibility has become a central concern in modern software engineering, not only as a matter of social responsibility but as a determinant of software quality and user experience~\cite{A1daThay}. Yet, despite growing awareness and regulatory pressure, many developers continue to struggle to design truly accessible solutions in corporate environments~\cite{WhyNotAccessible, GrundyPractitioners}. Accessibility is too often treated as a compliance checklist — something to be achieved at the end of the development cycle — rather than as a catalyst for innovation and creativity~\cite{PractitionersSpringer}. As a result, accessibility efforts frequently fail to produce sustainable cultural and technical change within software organizations~\cite{mckinsey2023diversity}.

Prior research has focused mainly on accessibility guidelines, automated testing tools, and general developer awareness~\cite{AccessbAndroid, JonJon}. However, Human-Computer Interaction (HCI) researchers highlight that a significant barrier to accessible software is the lack of knowledge and training among non-disabled developers, leading to poor application of accessibility standards in practice~\cite{FreireAcessibMob}. These works tend to frame accessibility as a technical problem to be solved rather than as a participatory co-creation process~\cite{EffectiveInclusion}.

What remains underexplored is how People with Disabilities (PWD) can act not only as beneficiaries but as drivers of accessible innovation, shaping and leading how accessibility is conceived within software teams. Moreover, the growing integration of Large Language Models (LLMs) into software projects introduces new challenges and opportunities: these models can enhance personalization, automate interface adaptation, and reduce cognitive barriers, but they can also encode biases that reproduce exclusion if accessibility is not considered from the outset~\cite{forgotten}.

This research aims to address this gap by investigating how people with and without disabilities collaborate to turn accessibility awareness into concrete innovation practices in LLM-assisted software projects.

The following research questions guided our study:
\begin{itemize}
    \item \textbf{RQ1}: What main accessibility pain points are addressed by industry practitioners? \\ \textbf{Rationale}: Understanding the concrete accessibility pain points experienced by developers and users with disabilities is essential to uncover how accessibility challenges are perceived and prioritized in real software projects.
    \item \textbf{RQ2}: How does the participation or leadership of PWD influence the direction or outcomes of the LLM-assisted software solutions? \\ \textbf{Rationale}: This perspective focuses on how PWD involvement in co-creating processes shapes problem framing, technical decisions, and the overall solution direction. By examining this dynamic, we explore accessibility not as a downstream validation step, but as a generative process driven by lived experience.
    \item \textbf{RQ3}: What is the role of LLMs in enhancing the development of innovative accessibility solutions? \\ \textbf{Rationale}: We sought to expand the comprehension of how LLMs are being appropriated by developers (with and without disabilities) to implement solutions that promote accessibility beyond traditional compliance-driven approaches.
\end{itemize}

To answer these questions, we analyzed all 14 submissions to \textit{Code Without Barriers}, an internal campaign led by a major Brazilian technology company (with around 3,000 employees) that aimed to foster the use of LLMs to design accessibility solutions.
In addition, we conducted a focus group discussion with nine participants to deepen the debate about their experiences in developing innovative solutions to promote accessibility.

This paper contributes to expanding empirical knowledge about the use of LLMs in developing accessibility solutions as well as the impact of PWD protagonism in leading and developing these solutions. In this regard, we observed how accessibility challenges were reframed as creative opportunities, particularly when PWD took leading roles in defining the technical and conceptual directions of the proposed solutions. Our research's main findings indicate that collaboration in mixed-ability teams fosters solutions to real-world accessibility issues, and that PWD perceive LLMs as a powerful accessibility ally.

\section{Related Work}

Despite the growing body of work on accessibility guidelines and tools, little is known about how developers with disabilities actively participate in shaping accessibility solutions within software development teams~\cite{EffectiveInclusion}.

Prior studies have investigated how AI-based systems can enhance accessibility, reveal embedded biases, and support inclusion for people with disabilities. Below, we summarize key contributions that inform and contrast with our investigation of accessibility-driven innovation in LLM-based projects.

He et al.~\cite{he2025gena11y} proposed \textit{GenA11y}, an automated tool that leverages LLMs to detect accessibility issues on the web. The authors identify limitations on traditional rule-based accessibility checkers, such as WAVE, which cover only a fraction of the WCAG success criteria. Using WCAG 2.2 as a foundation, the study classifies success criteria based on their testability and demonstrates how generative AI can interpret the semantic context of web elements to identify issues beyond the reach of static analysis tools. Results show that GenA11y detects, on average, eight additional accessibility violations per webpage compared to existing tools, underscoring the potential of LLMs to complement human judgment in accessibility verification.

\cite{gadiraju2023disability} examine how LLM perpetuate subtle forms of ableism and bias in their responses. The study is grounded in a disability-centered framework and conducted 19 focus groups with 56 participants with diverse disabilities. Through qualitative coding and co-annotation, participants analyzed chatbot responses to disability-related prompts, identifying patterns of harm.
The methodology centers people with disabilities as experts in identifying bias, and the results led to a taxonomy of harm that extends traditional notions of toxicity and offensiveness to include representational harm and stereotyping. The paper argues for participatory annotation practices and inclusive dataset design to mitigate bias in AI systems, proposing co-design between ML developers and disabled communities as a path toward fairer models.

\cite{mullen2024diary} investigate real-world use cases of LLM-based chatbots among people with disabilities. The study builds on the Disability Justice principle of ``\textit{following the leadership of the most impacted}''. Participants reported using chatbots for diverse purposes, including social reflection, writing assistance, emotional regulation, and leisure activities such as storytelling and games. The qualitative analysis identified nine use case categories, highlighting both empowerment opportunities and ongoing accessibility challenges. The authors conclude that LLM-based chatbots can foster autonomy and inclusion when designed with accessibility affordances in mind, but caution that their effectiveness depends on the extent to which users can modify and contextualize AI outputs to fit their specific needs.

Research has also explored the role of AI-mediated tools in supporting social interaction and communication.~\cite{kong2025interdependence} explore how chatbots can mediate social interactions between neurodivergent and neurotypical individuals. Through a 5-day diary study and semi-structured interviews with 16 participants (eight neurotypical and eight with intellectual disabilities), the authors identify barriers to mutual understanding—such as differing social expectations and communication uncertainties—and assess how chatbot-based social support could bridge these gaps. The methodology emphasizes bidirectional support rather than deficit correction, aligning with the interdependence and double empathy frameworks. Findings reveal that neurodivergent participants value chatbots as empathetic companions offering emotional support, while neurotypical participants prefer practical communication coaching. The study contributes design principles for chatbots that promote interdependence and reciprocal social support, reframing assistive technology as a means of mutual adaptation rather than normalization.

While these studies collectively advance understanding of accessibility and disability-centered design in the era of generative AI, they primarily focus on the interaction between individuals and technology — either in detecting accessibility issues, exposing algorithmic ableism, or designing more inclusive conversational agents. In contrast, our work investigates accessibility as a \textit{collaborative innovation process} within an organizational context, examining how PWDs actively participate and lead accessibility initiatives in LLM-based projects. Rather than centering solely on AI systems or user–AI interactions, our study foregrounds the socio-technical dynamics of co-creation, exploring how inclusive participation reshapes both technological outcomes and organizational culture toward accessibility-driven innovation.

\section{The \textit{Code Without Barriers} Program}

The \textit{Code Without Barriers} program was an internal corporate initiative, planned and coordinated by the Diversity and Inclusion area, that aimed to integrate inclusion and accessibility as essential dimensions of technological excellence. It emerged within a broader organizational effort to rethink how innovation is conceived and practiced—moving away from a compliance-oriented model toward a culture that recognizes accessibility and diversity as enablers of creativity, collaboration, and sustainable value creation.

The program sought to foster an organizational mindset that positions inclusion not as an external requirement but as a strategic and creative dimension of software development. By actively encouraging employees from diverse backgrounds and experiences—including People with Disabilities (PWD)—to contribute to the ideation and implementation of accessible solutions.
An important driver of this initiative was the stimulus to use the company's internal LLM-based infrastructure, the \textit{Stackspot AI}\footnote{https://ai.stackspot.com/}.

Importantly, the program created cross-functional spaces where technical specialists, designers, and managers — some of them PWD — could collaborate on shared accessibility challenges. Through these interactions, \textit{Code Without Barriers} helped transform individual awareness into collective learning. Rather than prescribing solutions, it functioned as an opportunity for experimentation and reflection, illustrating that accessibility, when framed as a creative and collaborative endeavor, can generate not only better products but also meaningful organizational learning and cultural evolution.

\section{Methodology}

This study adopted a qualitative research design organized in two main stages, combining document analysis and focus group discussions. The goal was to understand how people with and without disabilities engage in accessibility-driven innovation processes within corporate environments. Both stages considered the ideas proposed by participants in \textit{Code Without Barriers} campaign.

\subsection{Stage 1: Collection and Curation of Accessibility Proposals}

In the first stage, we collected written descriptions of accessibility-oriented solutions proposed. Participants were invited to submit their ideas through a dedicated post on the organization's intranet platform, allowing others to view, react, and provide feedback. The ideas could describe improvements to existing deliveries, inclusive automations, accessibility best-practice guides, or other accessibility-oriented initiatives, using LLM-based technology.

To guide submissions, the company provided a standardized template encouraging participants to reflect on both the technical and social dimensions of accessibility. Each submitted idea was expected to follow this structure:

\begin{enumerate}
    \item[1.] What accessibility problem does your solution aim to address?
    \item[2.] What is being created, adjusted, or presented?
    \item[3.] How can your idea generate a positive impact for PWD?
    \item[4.] What results do you expect/already achieved with the solution?
\end{enumerate}

After the first analysis, researchers requested by email more information about the submissions to better understand: PWD involvement in the solution's ideation, design, or implementation; and which Large Language Model(s) or AI technologies were applied.

During this review, it became evident that participants contributed at varying levels of maturity: while some focused primarily on the ideation of accessibility, oriented concepts, others advanced further to develop functional prototypes that demonstrated the practical feasibility of their ideas.

In practice, participants employed \textit{Stackspot AI} through two main mechanisms:
\textit{(a)} prompt-based commands, referred to internally as \textbf{Quick Commands}, which allowed developers to run short, context-aware interactions with an embedded model to automate documentation, text simplification, or code validation; and
\textit{(b)} custom-built \textbf{AI Agents}, which extended these prompts into reusable workflows that combined specification generation, cognitive assistance, and accessibility verification.

These LLM integrations were not prescribed by the research team; rather, they emerged organically from participants' daily work practices. As a result, the submissions provided a unique opportunity to observe how developers, both with and without disabilities, appropriated LLMs as creative tools to prototype accessible and inclusive solutions. This also established the empirical foundation for exploring RQ3, which investigates the role of LLMs as mediators in accessibility innovation.

\subsection{Stage 2: Focus Group Discussions}

In the second stage, the proponents of the 14 submitted proposals were invited to participate in a focus group. A focus group is a methodology that explores participants' experiences and perspectives through moderated group discussion, capturing both individual views and collective meaning-making~\cite{de2015using, hoda2024qualitative}. This stage aimed to deepen our understanding of how accessibility innovation emerges and evolves when PWDs actively participate or assume leadership roles in the definition, design, and implementation of inclusive software solutions.

The focus group used a semi-structured format, balancing consistency across topics with flexibility for participants to elaborate on their experiences. The discussion guide was designed to explore not only the social and organizational aspects of accessibility innovation but also the technical implications of incorporating emerging technologies such as LLMs. Specifically, participants were invited to reflect on:

\begin{itemize}
    \item PWD participation:
    \begin{itemize}
        \item How did their participation or leadership in defining and proposing accessibility solutions influence the solution itself?
        \item How did their participation help identify the problem and shape the construction of the solution?
    \end{itemize}
    \item AI contribution:
    \begin{itemize}
        \item What was the role of LLMs — particularly the \textit{Stackspot AI} platform — in enhancing the development of accessibility solutions?
        \item How can AI contribute to making software development more accessible, both for developers with disabilities and for end users?
    \end{itemize}
\end{itemize}

The focus group was facilitated by the first author, with co-moderation from the second author and the two company diversity area members who envisioned the initiative. All facilitators had prior experience with qualitative data collection, accessibility research, or accessibility in the workplace implementation.
The session was conducted online via Google Meet, enabling participants from different regions to join remotely. The meeting was recorded and automatically transcribed using Google Meet's transcription feature to ensure accurate data capture.
Participation was voluntary, and all attendees provided informed consent before the session began. From the 14 invited, a total of nine participants attended the focus group, among them, five who identified as People with Disabilities (two autistic, one hearing impaired, two blind).

One of the invited participants, who is deaf, was unable to attend the main focus group session because a Brazilian Sign Language (Libras) interpreter was unavailable at that time. To ensure inclusion, the first author and one Diversity and Inclusion Team member subsequently conducted an individual interview with this participant, accompanied by a qualified interpreter, using the same set of guiding questions as in the group discussion. The participants' responses were anonymized and later integrated into the focus group's overall dataset for analysis.

The proponents represented a mix of roles, including software developers, testers, designers, tech writers, and accessibility advocates. The discussion lasted approximately 60 minutes; 30 minutes for each question. The discussion produced rich qualitative data on participants' perceptions, experiences, and reflections regarding accessibility-driven innovation in LLM-based projects.

This research as part of a wider research project, and it was approved by the Ethics Committee of the Federal University of Pará under the number 74659423.6.0000.0018.

\subsection{Data Analysis}

The qualitative data collected from both stages — the written proposals (Stage 1) and focus group discussions (Stage 2) — were then analyzed using thematic analysis. We first conducted open coding on the proposals to identify the specific accessibility challenges (RQ1) and the LLM technologies used. We then performed focused coding on the focus group transcripts to analyze patterns related to PWD participation (RQ2) and leadership influence (RQ3), using a constant comparative method to ensure our findings were grounded in the participants' lived experiences and proposals. The first and second authors conducted coding and data analysis, and the last author revised the final codes.

\section{Results}
This section presents the main findings from the analysis of the 14 innovation proposals and supplementary focus group data. Table~\ref{tab:tabproposals} lists the submitted proposals and a brief description, highlighting the primary accessibility barriers addressed and the involvement of PWD in the design process. More information about all the accessibility innovation proposals, RQ1 codes, PWD involvement, LLM/AI used, and expected impact is detailed in the supplementary material\footnote{https://github.com/human-interaction-with-tecnologies/CodeWithoutBarriers}.

\begin{table}[]
\caption{Submitted Proposals}
\label{tab:tabproposals}
\resizebox{\textwidth}{!}{%
\begin{tabular}{l|l|l}
\textbf{Id} & \textbf{Proposal}                           & \textbf{Brief Description}                                        \\ \hline
P1          & Facilita Spec                               & Accessibility specs for prototypes aligned with WCAG.   \\ \hline
P2          & Design2Code 3.0                             & Embeds accessibility in code generated from Figma.                \\ \hline
P3          & Chrome Plugin for Image Description         & Captures webpages and returns contextual image descriptions.      \\ \hline
P4          & WhatsApp Accessibility Bot                  & Transcribes images/stickers in chats to accessible text.          \\ \hline
P5          & Image Reader Agent                          & Generates automatic alt text for images, reducing cognitive load. \\ \hline
P6          & Task \& Learning Organizer                  & Summarizes and structures study tasks to support comprehension.   \\ \hline
P7          & Marketing Agents with Inclusive Language    & Integrates inclusive and gender-neutral language in AI agents.    \\ \hline
P8          & AI-Supported Learning Method                & Customizes AI-assisted learning to reduce barriers in upskilling. \\ \hline
P9          & Automated Dev for Motor Accessibility       & Automates code/screens/tests to minimize physical effort.         \\ \hline
P10         & NVDA Add-on -- Image Describer              & Adds shortcut for instant image description in screen reader.     \\ \hline
P11         & Accessibility Verifier Agent                & Checks code against WCAG 2.2 and suggests fixes.                  \\ \hline
P12         & Text Simplifier for Cognitive Accessibility & Simplifies complex texts for clearer comprehension.               \\ \hline
P13         & Alt-Text Booster                            & Detects missing alt text and auto-generates descriptions.         \\ \hline
P14         & Texto Claro                                 & Text adaptation for better comprehension among deaf and hearing.  \\ \hline
\end{tabular}%
}
\end{table}

The subsequent analysis explores how these proposals reflect broader socio-technical dynamics observed during the focus group discussions. Specifically, the thematic analysis revealed three interconnected patterns that correspond directly to our research questions:
how accessibility pain points were identified and reframed as creative opportunities (RQ1);
how PWD participation and leadership shaped design directions (RQ2);
and how LLMs acted as mediators in the creation of innovative accessibility solutions (RQ3).

The results of our analysis are organized according the research questions,
representing real industry application of LLM as accessibility tool, expanding its role from a simple compliance task into a driver for innovation and inclusion.

\subsection{RQ1: Main accessibility pain points addressed.}
\noindent
\textbf{Communication Barriers.}
A variety of communication barriers were addressed in half of the analysed proposals. Image description solutions were coded this way since they collaborate to improve communication among blind and non-blind people, especially in virtual spaces, such as shared prototypes, WhatsApp groups, and social media posts. The difficulties deaf and hearing individuals face in standard Portuguese written communication were also stated as a usual source of misunderstandings and misinterpretations. Autistic developers mentioned communication barriers through difficulties in understanding complex contexts and sometimes long written messages. P14 author stated: ``\textit{I am deaf, and `Texto Claro' has helped me a lot in my daily life. When I write texts the way deaf people usually write, the tool corrects and adjusts the Portuguese, making the text easier and clearer to understand for hearing people. This improves my communication and facilitates inclusion.}''

\noindent
\textbf{Low Productivity.}
A motor-impaired developer mentioned being impacted in his productivity due to his physical limitations (using only one hand to type). Extra work performed by blind users to access visual information in pictures or software screens was also cited as an important pain point that deserved special attention from our blind participants. One of the P11 authors, who is a blind woman, mentioned: ``\textit{when we see these possibilities, for example, describing the screen like this and so quickly, right? Because like, ah, you can, but you have to download, take a screenshot, upload, describe, go back, validate. So, man, there are a lot of steps. And this feature, I think, also comes to contribute in this sense, you know?}''

\noindent
\textbf{Lack of Autonomy.}
Dependence on non-impaired colleagues to perform activities was expressed as a common pain point, an important incentive to propose solutions that make the workplace environment more accessible.
Blind developers and testers mentioned the need for constant assistance to generate and validate front-end code more independently. A blind tester said: ``\textit{I know someone who works here, who works as a functional (test analyst) and he goes through some difficulties, right, quite a lot in that sense, you know, of actually testing, of needing a friendly eye to see the context of the screen}''.

\noindent
\textbf{Self-organization and Learning issues.}
Specially (not exclusively) autistic people cited some difficulties in self-organization, time management, and learning new technologies. The P5/P6 proponent mentioned the creation of an: ``\textit{agent that summarizes study topics for me in a way that I can understand more easily}''. P9 proponent also mentioned ``\textit{I had difficulty learning things}''.

\noindent
\textbf{Lack of Accessibility.}
Tools that automate accessibility specification and verification during the software development phase were proposed to guarantee WCAG alignment. P1's proponent illustrated the accessibility issue, describing the accessibility specification process: ``\textit{(the process) is often manual, time-consuming, and prone to failure due to a lack of technical expertise or time constraints. This results in inconsistent or missing specifications, negatively impacting the experience of people with disabilities.}''.

\subsection{RQ2: Influence of the participation or leadership of PWD.}
Innovation in accessibility was most visible when PWD acted not as end-users, but as makers and creative agents within the development process. Their lived experience was not a constraint but a source of design insight and problem definition.

An example was \textit{Design2Code 3.0 (P2)}, co-validated by a blind developer, who expanded the scope of front-end generation to ensure semantic accuracy for screen readers. Instead of perceiving accessibility as post-hoc compliance, teams redefined it as a creative extension of product design. Similarly, in \textit{Marketing Agents with Inclusive Language (P7)}, neurodivergent contributors guided the tone and vocabulary of LLM-based communication agents, embedding inclusive language rules into the system's prompts.

PWD also played a critical role in testing and validating the solutions, ensuring their effectiveness and identifying areas for improvement. This hands-on involvement prevented hallucinations or solutions that did not truly meet the users' needs.
The importance of PWD's participation in accessibility solutions' definition can be illustrated in P10's proponent speech: ``\textit{Well, and then there was a moment that was really interesting, and afterward we were laughing about it, because one of the people on the team created the Add-on so that it would describe the image that the mouse pointer was over. And then it was cool because the (blind developer) very politely said: `Wow, very good, thank you, but we don't use the mouse'.}''.

\subsection{RQ3: LLMs role in accessibility solutions.}
In practice, participants employed \textit{Stackspot AI} through two main mechanisms:
\textit{(a)} prompt-based commands, referred to internally as \textbf{Quick Commands}, which allowed developers to run short, context-aware interactions with an embedded model to automate documentation, text simplification, or code validation; and
\textit{(b)} custom-built \textbf{AI Agents}, which extended these prompts into reusable workflows that combined specification generation, cognitive assistance, and accessibility verification.

Across multiple cases, LLM-based tools served as catalysts that expanded both cognitive and physical independence. Rather than replacing human action, they amplified users' capabilities—particularly for neurodivergent and blind developers—by reducing cognitive overload and enabling autonomy in technical tasks.

Participants with neurodivergent profiles emphasized that tools like the \textit{Image Reader Agent} and \textit{Facilita Spec} minimized the mental fatigue involved in repetitive writing and documentation. Blind developers, in turn, reported that the \textit{NVDA Add-on} and \textit{Accessibility Verifier} enabled faster navigation, on-demand image interpretation, and real-time accessibility validation, reducing dependency on external support.

While many of the initial ideas were born from individual needs—such as automating daily tasks, simplifying study routines, or describing images—these prototypes gradually converged into shared practices that fostered a broader sense of accessibility awareness within the organization.
This transformation was reflected in both the proposal dataset and focus group discussions, where accessibility became understood not as a peripheral obligation, but as part of the company's innovation mindset.

This gradual transition from individual experimentation to shared organizational practices indicates an emerging cultural shift—one in which accessibility operates simultaneously as a technical discipline and a collaborative value. Rather than constituting a whole ``movement'', these changes suggest the early stages of cultural consolidation, in which curiosity, empathy, and technological fluency sustain accessibility as an evolving practice supported by AI-mediated practices.

\section{Limitations}

As with most qualitative research, our study is subject to several limitations that should be acknowledged when interpreting the results.
First, the findings are derived from a single organizational context, which may limit the generalizability of the insights to other companies or cultural settings.
The \textit{Code Without Barriers} initiative reflects a specific corporate environment and accessibility maturity level, and different organizations may face distinct challenges or employ alternative mechanisms for inclusive innovation.

Although we analyzed both group and individual discussions (including an additional interview with a deaf participant and an interpreter), the asynchronous, mediated nature of these interactions may have influenced participants' comfort levels and the depth of their responses. In addition, the sample size was relatively small and self-selected, potentially attracting participants who were already engaged or motivated by accessibility topics.
Accessibility itself also shaped the research process—certain limitations in language interpretation and platform accessibility may have constrained participation for some individuals.

Finally, our analysis focused on participants' reflections and experiences rather than longitudinal behavioral changes or measurable project outcomes. Future studies could build on this work by conducting longitudinal or multi-organizational analyses to evaluate how participatory accessibility practices evolve and how they influence software development processes and products.

\section{Conclusions}

This study examined how people with and without disabilities collaborate to transform awareness of accessibility into innovation practices in LLM-assisted software projects. By analyzing 14 proposals and a follow-up focus group from a Brazilian company initiative, we found that accessibility evolves when reframed as a creative and participatory process rather than a compliance obligation. Our findings showed that accessibility pain points (RQ1) were reinterpreted as opportunities for creative problem-solving; that the participation and leadership of PWD (RQ2) shaped both design and direction of innovation; and that LLMs acted as mediators that enhanced cognitive and physical independence (RQ3).

Beyond individual achievements, these dynamics revealed early signs of an organizational learning process, supporting the position that generative AI, when used creatively, can extend rather than replace human agency, fostering~\textit{AI-augmented independence}. However, as Jackson et al.~\cite{GenAIInSoftDev2025} note, future research should consider continue to monitor how such tools affect developers' creativity, well-being, and sense of purpose, as well as how team structures and educational programs adapt to these shifts in software work.

From a socio-technical perspective, these findings echo calls for more critical and liberative approaches to accessible technology. Rocha et al.~\cite{EffectiveInclusion} emphasize that authentic inclusion of PWD in software development transforms not only products but also professional practices. Similarly, disability scholars have argued for ``\textit{counterventions}''—technological interventions that challenge normative assumptions about disability and expertise~\cite{RuaCounterventions}. In this sense, the practices observed here offer a glimpse of counter-hegemonic accessibility: one that empowers PWD as innovators, redefines participation as leadership, and uses AI not to normalize difference but to design with it.

In practical terms, our results suggest that corporate AI ecosystems should integrate accessibility as a creative discipline supported by participatory and critical design practices. Organizations are encouraged to create spaces where PWD can lead co-design with LLMs, embedding inclusive thinking into everyday development. Future work should longitudinally explore how these inclusive, AI-mediated collaborations evolve across educational, industrial, and civic contexts—and how they can inform a new generation of accessibility research grounded in creativity and interdependence.

\section*{Acknowledgments}
The authors thank the Zup Innovation Diversity and Inclusion Team, on behalf of Ana Duarte and Mariana Silva, for planning and executing this initiative, and for receiving our research as part of it.
This study was partially funded by CAPES - Finance Code
001, INES.IA (National Institute of Science and Technology for Software Engineering Based on and for Artificial Intelligence, \url{www.ines.org.br}), and CNPq (408817/2024-0, 314680/2026-8 and 308623/2022-3).

\bibliographystyle{unsrt}
\bibliography{referencias}

@article{A1daThay,
author = {Rocha, Thayssa Aguila da and Teran, Luciano Arruda and Melo, Giselle Lorrane Nobre and Menezes, Nicoly da Silva and da Silva, Ingrid Moreira Miranda and de Souza, Cleidson Ronald Botelho and Mota, Marcelle Pereira},
title = {How are People with Disabilities Embraced in Software Development Teams? A Systematic Literature Review},
year = {2025},
issue_date = {November 2025},
publisher = {Association for Computing Machinery},
address = {New York, NY, USA},
volume = {9},
number = {7},
url = {https://doi.org/10.1145/3757610},
doi = {10.1145/3757610},
journal = {Proc. ACM Hum.-Comput. Interact.},
month = oct,
articleno = {CSCW429},
numpages = {29}
}

@article{GrundyPractitioners,
author = {Bi, Tingting and Xia, Xin and Lo, David and Grundy, John and Zimmermann, Thomas and Ford, Denae},
title = {Accessibility in Software Practice: A Practitioner’s Perspective},
year = {2022},
issue_date = {October 2022},
publisher = {Association for Computing Machinery},
address = {New York, NY, USA},
volume = {31},
number = {4},
issn = {1049-331X},
url = {https://doi.org/10.1145/3503508},
doi = {10.1145/3503508},
journal = {ACM Trans. Softw. Eng. Methodol.},
month = jul,
articleno = {66},
numpages = {26}
}

@article{PractitionersSpringer,
author = {Seixas Pereira, Leticia and Duarte, Carlos},
title = {Evaluating and monitoring digital accessibility: practitioners’ perspectives on challenges and opportunities},
year = {2025},
issue_date = {August 2025},
publisher = {Association for Computing Machinery},
address = {New York, NY, USA},
volume = {24},
number = {3},
issn = {1615-5297},
url = {https://doi.org/10.1007/s10209-025-01210-w},
doi = {10.1007/s10209-025-01210-w},
journal = {Universal Access in the Information Society},
month = ago,
numpages = {18},
}

@article{FreireAcessibMob,
title = {Accessibility in the mobile development industry in Brazil: Awareness, knowledge, adoption, motivations and barriers},
journal = {Journal of Systems and Software},
volume = {177},
pages = {110942},
year = {2021},
issn = {0164-1212},
doi = {https://doi.org/10.1016/j.jss.2021.110942},
url = {https://www.sciencedirect.com/science/article/pii/S016412122100039X},
author = {Manoel Victor Rodrigues Leite and Lilian Passos Scatalon and André Pimenta Freire and Marcelo Medeiros Eler},
}

@article{GenAIInSoftDev2025,
author = {Jackson, Victoria and Vasilescu, Bogdan and Russo, Daniel and Ralph, Paul and Prikladnicki, Rafael and Izadi, Maliheh and D’Angelo, Sarah and Inman, Sarah and Andrade, Anielle and van der Hoek, Andr\'{e}},
title = {The Impact of Generative AI on Creativity in Software Development: A Research Agenda},
year = {2025},
issue_date = {June 2025},
publisher = {Association for Computing Machinery},
address = {New York, NY, USA},
volume = {34},
number = {5},
issn = {1049-331X},
url = {https://doi.org/10.1145/3708523},
doi = {10.1145/3708523},
journal = {ACM Trans. Softw. Eng. Methodol.},
month = may,
articleno = {133},
numpages = {28}
}

@inproceedings{AccessbAndroid,
author = {Alshayban, Abdulaziz and Ahmed, Iftekhar and Malek, Sam},
title = {Accessibility issues in Android apps: state of affairs, sentiments, and ways forward},
year = {2020},
isbn = {9781450371216},
publisher = {Association for Computing Machinery},
address = {New York, NY, USA},
url = {https://doi.org/10.1145/3377811.3380392},
doi = {10.1145/3377811.3380392},
booktitle = {Proceedings of the ACM/IEEE 42nd International Conference on Software Engineering},
pages = {1323–1334},
numpages = {12},
location = {Seoul, South Korea},
series = {ICSE '20}
}

@inproceedings{JonJon,
author = {Mack, Kelly and McDonnell, Emma and Jain, Dhruv and Lu Wang, Lucy and E. Froehlich, Jon and Findlater, Leah},
title = {What Do We Mean by “Accessibility Research”? A Literature Survey of Accessibility Papers in CHI and ASSETS from 1994 to 2019},
year = {2021},
isbn = {9781450380966},
publisher = {Association for Computing Machinery},
address = {New York, NY, USA},
url = {https://doi.org/10.1145/3411764.3445412},
doi = {10.1145/3411764.3445412},
booktitle = {Proceedings of the 2021 CHI Conference on Human Factors in Computing Systems},
articleno = {371},
numpages = {18},
location = {Yokohama, Japan},
series = {CHI '21}
}

@inproceedings{WhyNotAccessible,
author = {Patel, Rohan and Breton, Pedro and Baker, Catherine M. and El-Glaly, Yasmine N. and Shinohara, Kristen},
title = {Why Software is Not Accessible: Technology Professionals' Perspectives and Challenges},
year = {2020},
isbn = {9781450368193},
publisher = {Association for Computing Machinery},
address = {New York, NY, USA},
url = {https://doi.org/10.1145/3334480.3383103},
doi = {10.1145/3334480.3383103},
booktitle = {Extended Abstracts of the 2020 CHI Conference on Human Factors in Computing Systems},
pages = {1–9},
numpages = {9},
location = {Honolulu, HI, USA},
series = {CHI EA '20}
}

@inproceedings{EffectiveInclusion,
author = {da Rocha, Thayssa A and de Souza, Cleidson and Teran, Luciano and Mota, Marcelle},
title = {Effective Inclusion of People with Disabilities in Software Development Teams},
year = {2024},
isbn = {9798400710476},
publisher = {Association for Computing Machinery},
address = {New York, NY, USA},
url = {https://doi.org/10.1145/3674805.3690749},
doi = {10.1145/3674805.3690749},
booktitle = {Proceedings of the 18th ACM/IEEE International Symposium on Empirical Software Engineering and Measurement},
pages = {447–453},
numpages = {7},
location = {Barcelona, Spain},
series = {ESEM '24}
}

@misc{mckinsey2023diversity,
  author       = {Dixon-Fyle, Sundiatu and Hunt, Vivian and Huber, Celia and Márquez, Maria del Mar Martinez and Prince, Sara and Thomas, Ashley},
  title        = {Diversity Matters Even More – The Case for Holistic Impact},
  year         = {2023},
  month        = nov,
  url          = {https://www.mckinsey.com},
  note         = {Accessed: 2025-10-23}
}

@inproceedings{RuaCounterventions,
author = {Williams, Rua Mae and Boyd, Louanne and Gilbert, Juan E.},
title = {Counterventions: a reparative reflection on interventionist HCI},
year = {2023},
isbn = {9781450394215},
publisher = {Association for Computing Machinery},
address = {New York, NY, USA},
url = {https://doi.org/10.1145/3544548.3581480},
doi = {10.1145/3544548.3581480},
booktitle = {Proceedings of the 2023 CHI Conference on Human Factors in Computing Systems},
articleno = {653},
numpages = {11},
location = {Hamburg, Germany},
series = {CHI '23}
}

@inproceedings{forgotten,
author = {Alshaigy, Bedour and Grande, Virginia},
title = {Forgotten Again: Addressing Accessibility Challenges of Generative AI Tools for People with Disabilities},
year = {2024},
isbn = {9798400709654},
publisher = {Association for Computing Machinery},
address = {New York, NY, USA},
url = {https://doi.org/10.1145/3677045.3685493},
doi = {10.1145/3677045.3685493},
booktitle = {Adjunct Proceedings of the 2024 Nordic Conference on Human-Computer Interaction},
articleno = {68},
numpages = {6},
location = {Uppsala, Sweden},
series = {NordiCHI '24 Adjunct}
}

@String{Computing = "Computing" }

@String{Springer = "Springer-Verlag" }

@inproceedings{mullen2024diary,
author = {Mullen, Kayla and Xue, Wenhan and Kudumu, Manasa},
title = {“I'm treating it kind of like a diary”: Characterizing How Users with Disabilities Use AI Chatbots},
year = {2024},
isbn = {9798400706776},
publisher = {Association for Computing Machinery},
address = {New York, NY, USA},
url = {https://doi.org/10.1145/3663548.3688549},
doi = {10.1145/3663548.3688549},
booktitle = {Proceedings of the 26th International ACM SIGACCESS Conference on Computers and Accessibility},
articleno = {133},
numpages = {7},
location = {St. John's, NL, Canada},
series = {ASSETS '24}
}

@article{he2025gena11y,
author = {He, Ziyao and Huq, Syed Fatiul and Malek, Sam},
title = {Enhancing Web Accessibility: Automated Detection of Issues with Generative AI},
year = {2025},
issue_date = {July 2025},
publisher = {Association for Computing Machinery},
address = {New York, NY, USA},
volume = {2},
number = {FSE},
url = {https://doi.org/10.1145/3729371},
doi = {10.1145/3729371},
journal = {Proc. ACM Softw. Eng.},
month = jun,
articleno = {FSE101},
numpages = {24}
}

@inproceedings{gadiraju2023disability,
author = {Gadiraju, Vinitha and Kane, Shaun and Dev, Sunipa and Taylor, Alex and Wang, Ding and Denton, Remi and Brewer, Robin},
title = {"I wouldn't say offensive but...": Disability-Centered Perspectives on Large Language Models},
year = {2023},
isbn = {9798400701924},
publisher = {Association for Computing Machinery},
address = {New York, NY, USA},
url = {https://doi.org/10.1145/3593013.3593989},
doi = {10.1145/3593013.3593989},
booktitle = {Proceedings of the 2023 ACM Conference on Fairness, Accountability, and Transparency},
pages = {205–216},
numpages = {12},
location = {Chicago, IL, USA},
series = {FAccT '23}
}

@inproceedings{kong2025interdependence,
author = {Kong, Ha-Kyung and Lowy, Rachel and Choi, Youjin and Kim, Jennifer G},
title = {Working Together Toward Interdependence: Chatbot-Based Support for Balanced Social Interactions Between Neurodivergent and Neurotypical Individuals},
year = {2025},
isbn = {9798400713941},
publisher = {Association for Computing Machinery},
address = {New York, NY, USA},
url = {https://doi.org/10.1145/3706598.3713344},
doi = {10.1145/3706598.3713344},
booktitle = {Proceedings of the 2025 CHI Conference on Human Factors in Computing Systems},
articleno = {779},
numpages = {17},
location = {
},
series = {CHI '25}
}

@article{hoda2024qualitative,
  title={Qualitative research with socio-technical grounded theory},
  author={Hoda, Rashina},
  journal={Springer},
  year={2024},
  publisher={Springer}
}

@inproceedings{de2015using,
  title={Using Focus Group in Software Engineering: lessons learned on characterizing software technologies in academia and industry.},
  author={de Fran{\c{c}}a, Breno Bernard Nicolau and Ribeiro, Talita Vieira and dos Santos, Paulo S{\'e}rgio Medeiros and Travassos, Guilherme Horta},
  booktitle={CIbSE},
  pages={351},
  year={2015}
}

\end{document}